\documentclass[12pt,a4paper]{article}%
\usepackage{amsmath}%
\usepackage{amsfonts}%
\usepackage{amssymb}%
\usepackage{graphicx}
\providecommand{\U}[1]{\protect\rule{.1in}{.1in}}
\newtheorem{theorem}{Theorem}
\newtheorem{acknowledgement}[theorem]{Acknowledgement}

\allowdisplaybreaks

\begin{document}
\clearpage

\begin{center}
Massive Gravity with $N=1$ Local Supersymmetry
\end{center}
\vspace{0.5cm}
\begin{center}
O. Malaeb \\
Physics Department, American University of Beirut, Lebanon
\end{center}
\vspace{3cm}
%

\begin{abstract}
A consistent theory of massive gravity, where the graviton acquires mass by spontaneously breaking diffeomorphism invariance, is now well established. We supersymmetrize this construction using $N=1$ fields. Coupling to $N=1$ supergravity is done by applying the rules of tensor calculus to construct an action invariant under local $N=1$ supersymmetry. The supersymmetric action is shown, at the quadratic level, to be free of ghosts and have as its spectrum a massive graviton, two gravitinos (with different masses) and a massive vector.
\end{abstract}

\clearpage

In 1970, it was concluded by van Dam, Veltman and Zakharov \cite{vandam}, \cite{zakharov} that the graviton mass must be rigorously zero. This was so because they noticed that in the linearized gravitation theory there is no smooth limit with respect to the graviton mass. The action they used was that of Fierz and Pauli \cite{fierzpauli} with mass terms breaking general coordinate invariance. Comparison of massive and massless theories with experiment, in particular the perihelion movement of Mercury, gives results that are different than that of General Relativity. So, this sets the graviton mass to be mathematically zero. This effect was first thought to be a no-go theorem for massive theories of gravity and was known as the van Dam-Veltman-Zakharov (vDVZ) discontinuity \cite{zakharov}, \cite{deser}. However, Vainshtein showed that the perturbation theory breaks down for the consideration of zero mass limit, therefore the discontinuity could be an artifact \cite{vainshtein}. The continuous in mass solution is constructed beyond the perturbation theory approach. So, a small non-zero graviton mass is no more excluded and does not contradict experiments. However, the massive gravity was still considered an ill-behaved theory since Deser and Boulware \cite{deser} showed that in the massive theory the extra scalar degree of freedom reappears at nonlinear level and does not decouple. Further developments were considered. Isham, Salam and Strathdee \cite{iss} examined a theory of bigravity. Then Chamseddine, Salam and Strathdee \cite{css} generalized their theory. While Dvali, Gabadadze, and Porrati \cite{dgp} considered theories with higher dimensions. \newline
Arkani-Hamed, Georgi and Schwartz \cite{ahgs} used four scalars in massive gravity to get a diffeomorphism invariant model. The idea of introducing four scalars was presented before by Siegel \cite{siegel}. 't Hooft \cite{hooft} (see also \cite{kaku}) considered a collection of four scalar fields that break coordinate reparametrization invariance to give mass to the graviton. The kinetic energies of the scalar fields involve a ghost in the unbroken phase. In the broken symmetry phase, there is no Fierz-Pauli term for the massive graviton, and the ghost state could not be decoupled.\newline
Chamseddine and Mukhanov \cite{simpleandelegant} showed how to make the graviton massive via Higgs mechanism. They employed four scalars with global Lorentz symmetry. As a result of symmetry breaking, a graviton absorbs all scalars and become massive spin 2 particle with five degrees of freedom. The resulting theory is unitary and free of ghosts. In \cite{puzzle}, they considered the massless limit of Higgs gravity and determined the Vainshtein scale. It was shown that below this scale, General Relativity is restored. Then, a simplified formulation of massive gravity where the Higgs fields have quadratic kinetic term was presented \cite{quadratic}.\newline
In \cite{rhamgabadadze}, massive gravity theories were constructed where it was shown that the Boulware-Deser ghost was avoided to all orders. This was followed by a covariant non-linear formulation of massive gravity which excludes the ghost in the decoupling limit \cite{rhametal}. In \cite{hassanrosen} and \cite{hassanetal}, ghost-free non-linear massive gravity models were worked out for flat and curved fiducial metric. Recently, Deser and Waldron \cite{deserwaldron} showed that Wess-Zumino massive gravity theory, which is free of ghosts, is acausal. 
\paragraph{}
In this letter, we give the results of formulating a theory of massive supergravity where the graviton and the gravitino both get masses due to the breakdown of diffeomorphism invariance. To do this, we generalize the Higgs mechanism adopted to the consistent formulation of massive gravity. Supersymmetrizing massive gravity is of physical interest because at the end our results will contain not only a gravitino, but also a massive spin-$3/2$ particle. Therefore, our theory will be similar to $N = 2$ supergravity, where two spin-$3/2$ particles exist.
\paragraph{}
\noindent To formulate our theory, we generalize what was done in the bosonic case \cite{simpleandelegant}. For this, we start with a set of four chiral superfields $\Phi^{A}\left( x,\theta,\overline{\theta}\right)$ subject to the conditions%
\[
\overline{D}_{\overset{.}{\alpha}}\Phi^{A}\left(  x,\theta,\overline{\theta}\right)  =0,
\]
where $A=0,1,2,3$ is a global Lorentz index. These chiral superfields are given by
\begin{equation}
\Phi_A = \varphi_A + i(\theta \sigma^{\mu} \bar{\theta})\partial_{\mu} \varphi_A - \frac{1}{4} \theta\theta \bar{\theta}\bar{\theta} \partial_{\mu}\partial^{\mu} \varphi_A + \sqrt{2} \theta \psi_A - \frac{i}{\sqrt{2}} \theta \theta \left(\partial_{\mu} \psi_A \sigma^{\mu} \bar{\theta}\right) + \theta\theta F_A.
\end{equation}
In the bosonic case \cite{simpleandelegant} (see also \cite{puzzle}), there is an induced metric $H^{AB}=g^{\mu\nu}\partial_{\mu}\varphi^{A}\partial_{\nu}\varphi^{B}$, and the action is at least quartic in the fields $\varphi^{A}$. There is a simpler quadratic action that could be taken through the use of auxiliary fields \cite{quadratic}, but this is less obvious to generalize to the supersymmetric case. It would be interesting to generalize the work done in \cite{rhametal} using this quadratic formulation, but that would need to use the superspace formulation of Wess and Bagger for supergravity \cite{wess}. This is a difficult take which we may revisit in the future. 
\paragraph{} 
We then define the basic field%
\begin{equation}
H_{ABC} = D^{\alpha} \Phi_A (\sigma_B)_{\alpha \dot{\alpha}} \bar{D}^{\dot{\alpha}} \Phi_C^{*} = D\Phi_A \sigma_B \bar{D} \Phi_C^*
\end{equation}
The Hermitian conjugate of this field is
\[
H_{ABC}^{\ast}=D\Phi_{C}\sigma_{B}\overline{D}\Phi_{A}^{\ast}=H_{CBA}%
\]
To simplify our expressions we denote $H_{ABC} \eta^{AB}$ by $H_{AAC}$. The D-type terms that could be formed from the products of this H field are given in the detailed paper \cite{ola}.\newline
Similar to the bosonic case \cite{simpleandelegant}, it is possible to avoid the complexity of adding the higher order terms by considering instead
\[
\overline{H}_{ABC}=H_{ABC}-Dx_{A}\sigma_{B}\overline{D}x_{C}^{\ast}%
\]
where $x_A$ are the coordinates. \newline
In addition, we can write F-type terms such as%
\begin{align}
{}& \overline{D}^{2}\left(  D\Phi_{A}\sigma^{AB}D\Phi_{B}\right), \quad \overline{D}^{2}\left(D\Phi_{A}D\Phi^{A} \bar{D}\Phi_{B}^{*} \bar{D}\Phi^{B*} \right).
\end{align}
\paragraph{}
To form our action, we must compose a matter action from the D-type and F-type terms and then couple their components to Supergravity by applying the rules of tensor calculus (check \cite{ola} for details). The rules of tensor calculus of chiral and vector multiplets are given in \cite{cremmer} (see also \cite{chamsbook}). \newline
The aim is to get an action after expanding the metric in terms of the vierbein, $g^{\mu \nu} = e^{\mu}_a e^{\nu a}$, and expanding the fields around the vacuum solution
\begin{equation}
\varphi^A = x^A + \chi^A, \quad \quad e^{\mu}_{a} = \delta^{\mu}_{a} + \bar{e}^{\mu}_{a},
\end{equation}
with
\begin{itemize}
\item Fierz-Pauli term for the graviton $(\bar{e}^{\mu}_A \bar{e}^A_{\mu} - \bar{e}^2)$
\item No linear terms for the vierbein
\item Maxwell term for the vectors 
\begin{equation}
l \left( \partial_{\mu}\chi_A \partial^{\mu} \chi^{A*} - \partial_A \chi^A \partial_B \chi^{B*}\right) 
\end{equation}
where $l$ is a constant
\item Free of ghosts; i.e. there shouldn't be terms like
\begin{equation} 
\partial_{\mu}\chi_A\partial^{\mu}\chi^A, \quad \text{or} \quad \partial_A \chi^A \partial_B \chi^B
\end{equation}
\item Mass terms for gravitinos
\end{itemize}
All these conditions, except the last, can be obtained by considering only D-type terms. Vector multiplets alone don't make the gravitino massive; therefore, chiral multiplets should be added because such terms give mass to the graviton.
\paragraph{}
Calculations showed that the matter action is composed of three vector multiplets and two scalar multiplets and is given by
\begin{align}
{}& m^4 \int \left(c_1 \bar{H}_{ABC} \bar{H}_{BCA} + c_2 \bar{H}_{ABB} \bar{H}_{CCA} + c_3 \bar{H}_{AB} \bar{H}_{AB}^*  \right)  d\theta^2 d\bar{\theta}^2 d^4x \nonumber \\
& + \frac{m^2}{\kappa} \int \left( c_4 \bar{D}^2\left(D\Phi_A \sigma^{AB} D\Phi_B\right) + c^{*}_4 {D}^2\left(\bar{D}\Phi^{*}_A \bar{\sigma}^{AB} \bar{D}\Phi_B^{*}\right) \right) d\theta^2 d^4x  \nonumber \\
& + m^4 \int c_5 \bar{D}^2\left(D\Phi_A D\Phi^A \bar{D} \Phi_B^{*} \bar{D} \Phi^{B *}\right) d\theta^2 d^4x \nonumber \\
& +m^4 \int  c_5^{*} D^{2}\left( \bar{D}\Phi_{A}^{*} \bar{D}\Phi^{A*} D\Phi_{B} D\Phi^{B} \right) d\theta^2 d^4x
\end{align}
where $H_{AB} = D\Phi_A  {D}\Phi_B$ and $m$ and $\kappa$ are used to fix the dimensions.\newline
The constants are determined by forcing the above mentioned constraints where details are given in \cite{ola}. At the end, the full Lagrangian $(e^{-1}L_{F} + e^{-1}L_{D})$ is given by
\begin{align}
{}& -\frac{1}{2} m^4 \ \left(\partial_{\mu} \chi_A \partial^{\mu} \chi^{A*} - \partial^{A} \chi_A \partial^{B} \chi_B^*\right) + \frac{7}{3} m^4 \ \left(\bar{e}^A_{\mu} \bar{e}^{\mu}_A - \bar{e}^2\right) - m^4 \ F_A F^{A *} \nonumber \\
& -\frac{7}{3} m^4 \ \left(\bar{e} \partial_A \chi^A +  \bar{e} \partial_A \chi^{A *}\right) + \frac{7}{3} m^4 \ \left(\bar{e}^{\mu}_A \partial_{\mu} \chi^A + \bar{e}^{\mu}_A \partial_{\mu} \chi^{A *}\right) \nonumber \\
& -\frac{5}{24} m^4 \ \epsilon^{ABCD} \bar{\psi}_A \gamma_B \gamma_5 \partial_C {\psi}_D + \frac{3i}{8} m^4 \ \bar{\psi}_A \gamma_{\mu} \partial^{\mu} {\psi}^A - \frac{\sqrt{6}}{8} m^6 \kappa \ \bar{\psi}_A \psi^A \nonumber \\
& + \frac{\sqrt{6}}{18} m^6 \kappa \ \bar{\psi}_A \gamma^A \gamma^B \psi_B - \frac{5 \sqrt{6}}{36} m^6 \kappa \ \bar{\psi}_A \gamma^B \gamma^A \psi_B + \frac{1}{2} e^{-1} \epsilon^{\mu \nu \rho \sigma} \bar{\phi}_{\mu} \gamma_5 \gamma_{\nu} \partial_{\rho} \phi_{\sigma} \nonumber \\
& + \frac{\sqrt{2} i}{4} m^4 \kappa \ \bar{\phi}_{\mu} \gamma^{\mu} \gamma^B \psi_B -  \frac{\sqrt{2} i}{4} m^4 \kappa \ \bar{\psi}_A \gamma^A \gamma^{\mu} \phi_{\mu} + \frac{\sqrt{3}}{6} m^2 \bar{\phi}_{\mu} \gamma^{\mu} \partial^A \psi_A  \nonumber \\
& + \frac{\sqrt{3}}{6} m^2 \partial^A \bar{\psi}_A \gamma^{\mu} {\phi}_{\mu} + \frac{\sqrt{6} i}{3} m^2 \kappa\ \bar{\phi}_{\mu} \gamma^{\mu \nu} \phi_{v} + \frac{\sqrt{3}}{12} m^2 \kappa \ \bar{\phi}_{\mu} \gamma^{\mu} \gamma^A \gamma^B \partial_B \psi_A \nonumber \\
& + \frac{\sqrt{3}}{12} m^2 \kappa \ \partial_B \bar{\psi}_A \gamma^{B} \gamma^A \gamma^{\mu} {\phi}_{\mu}  - \frac{1}{2 \kappa^2} R(e,w).
\end{align}
where we ignored terms higher than quadratic order. It can be seen that $m$ and $\kappa$ fixes the dimensions, where we have $[\chi_A] =-1, [\bar{e}]=0, [F_A]=0, [\psi_A] = -1/2$ and the gravitino $\phi_{\mu}$ has dimensions $3/2$.
\paragraph{} \noindent
From the action, the equations of motion of $\bar{\psi}_A$ and $\bar{\phi}_A$ are found. Then $\psi_A$ is decomposed 
\begin{align}
{}& \psi_{A} = \hat{\psi}_{A} + \frac{1}{4} \gamma_A \gamma_5 \lambda \nonumber \\
&\Rightarrow \bar{\psi}_{A} = \bar{\hat{\psi}}_{A} + \frac{1}{4} \bar{\lambda} \gamma_A \gamma_5,
\end{align}
where $\gamma_A \hat{\psi}^A = 0$. Similarly, we decomposed $\phi_{\mu}$
\begin{align}
{}& \phi_{\mu} = \hat{\phi}_{\mu} + \frac{1}{4} \gamma_{\mu} \gamma_5 \eta \nonumber \\
&\Rightarrow \bar{\phi}_{\mu} = \bar{\hat{\phi}}_{\mu} + \frac{1}{4} \bar{\eta} \gamma_{\mu} \gamma_5,
\end{align}
where $\gamma_A \hat{\phi}^A = 0$. In this decomposition, $\hat{\psi}_{A}$ and $\hat{\phi}_{\mu}$ are spin-$3/2$ helicities while $\lambda$ and $\eta$ are spin-$1/2$ helicities.\paragraph{}
The simplified equations of motion are found to be (for details, check \cite{ola})
\begin{align}
{}& \partial_A \hat{\psi}^A = \frac{3}{8} \gamma^5 \gamma^A \partial_A \lambda + \frac{9\sqrt{6} i}{8} m^2 \kappa \gamma^5 \lambda + 3 \sqrt{2} \kappa \gamma^5 \eta \nonumber \\
& \partial_A \hat{\phi}^A = \frac{-\sqrt{3}i}{2} m^2 \gamma^5 \gamma^A \partial_A \lambda + \frac{11\sqrt{2} }{2} m^4 \kappa \ \gamma^5 \lambda - \frac{3}{4} \gamma^5 \gamma_A \partial^A \eta - \frac{7 \sqrt{6} i}{2} m^2 \kappa \gamma^5 \eta \nonumber \\
& \text{and} \nonumber \\
& -\frac{5\sqrt{6}}{24} m^4 \kappa \ \gamma^A \partial_A \lambda - \frac{109 i}{4} m^6 \kappa^2 \ \lambda - 4\sqrt{2} i m^2 \kappa \ \gamma^{A} \partial_{A} \eta - \frac{137 \sqrt{3}}{6} m^2 \kappa^2 \ \eta \nonumber \\
& - 2\sqrt{3} \partial_A \partial^A \eta = 0.
\end{align}
Upon choosing the gauge $\eta = 0$, the equation of lambda is given by 
\begin{equation}
\gamma_A \partial^A \lambda + \frac{109 \sqrt{6}}{5} im^2 \kappa \lambda = 0
\end{equation}
This gives a Dirac type equation for the spin-1/2 helicities. 
\paragraph{}
Counting degrees of freedom, we have before coupling to supergravity a $N=1$ supersymmetry model (similar to Wess-Zumino model) having four spin-$0$ particles. $\varphi^0$ decouples due to Fierz-Pauli choice, therefore we have six (3 times 2) bosonic degrees of freedom. Also, we have six fermionic degrees of freedom forming a multiplet. This was coupled to supergravity which contains only one massless spin-2 graviton (two degrees of freedom) and one massless spin-$3/2$ gravitino (also having two degrees of freedom). Therefore, we started with an overall eight fermionic degrees of freedom and eight bosonic degrees of freedom. \newline
After coupling to supergravity, we get $N=1$ massive representation containing one massive spin-2 particle (five degrees of freedom), two massive spin-$3/2$ particles (four degrees of freedom each) and one massive vector field (spin-1 particle) with three degrees of freedom. Therefore, we have as a total eight fermionic and eight bosonic degrees of freedom. \newline
It should be noted that this is similar to the $N=2$ supersymmetry in which we have two massive spin-$3/2$ particles. However, the main difference is that in $N=2$ supersymmetry we have two gravitinos with same masses. In our case, supersymmetry is completely broken (at the same scale with diffeomorphism breaking) where different components split and thus we are left with two massive spin-$3/2$ particles with completely different masses. \paragraph{}

Summarizing, we gave a derivation of supersymmetrizing massive gravity. We generalized the Higgs mechanism used in the formulation of massive gravity to write down a massive supergravity action. To write the full Lagrangian, our analysis was carried in terms of the component fields of the supermultiplets, using the rules of tensor calculus for chiral and vector multiplets. At the end, the equations of motion were analyzed. We did not touch on the structure at the non-linear level. Analysis at higher orders may reveal ghosts, but this is for future work. 

\begin{acknowledgement}
I would like to thank Professor Ali Chamseddine for suggesting the problem and for his many helpful discussions in the subject.\newline
I would like to thank the American University of Beirut (Faculty of Science) for support.
\end{acknowledgement}

\end{document}